\documentclass[11pt,a4paper]{article}
\usepackage{amsmath,latexsym,fullpage,amssymb,color}
\usepackage{epic,eepic,ifthen,graphics,epsfig}
\usepackage[english]{babel}
\usepackage{times}
\usepackage{float}
\usepackage{xspace}
\usepackage[T1]{fontenc}
\newlength {\squarewidth}

\newcommand{\toto}{xxx}

\newcounter{linecounter}
\newcommand{\linenumbering}{\ifthenelse{\value{linecounter}<10}
{(\arabic{linecounter})}{(\arabic{linecounter})}}
\renewcommand{\line}[1]{\refstepcounter{linecounter}\label{#1}\linenumbering}

\renewcommand{\thelinecounter}{\ifnum \value{linecounter} > 
9 \else \fi\arabic{linecounter}}
\newfloat{algorithm}{thp}{lop}
\floatname{algorithm}{Algorithm}
\newfloat{construction}{thp}{lop}
\floatname{construction}{Construction}
\newcommand{\Xomit}[1]{}
\newcommand{\wwrite}{{\sf write}}  
\newcommand{\rread}{{\sf read}}  
\newcommand{\op}{{\sf op}} 
\newcommand{\push}{{\sf push}} 
\newcommand{\pop}{{\sf pop}}

\newcommand{\assign}{\mathit{Assignment}}
 
\newcommand{\done}{\tt{ok}} 
 
\newcommand{\vide}{{\bot}}
\newcommand{\cob}{\sf{co\_broadcast}} 
\newcommand{\wait}{\sf{wait}} 
\newcommand{\return}{\sf{return}}

\begin{document}

\title{\bf  Extending Causal Consistency  to\\
any Object Defined by a Sequential Specification}

\author{Achour Most\'efaoui$^{\dag}$,~ Matthieu Perrin$^{\circ}$\thanks{This 
work was done while this author was at LINA, Universit\'e de Nantes, France.},~
        Michel Raynal$^{\star,\ddag}$\\~\\
$^{\dag}$LINA, Universit\'e de Nantes, 44322 Nantes, France \\
        $^{\circ}$Computer Science Department, The Technion, Haifa, Isra\"{e}l \\
$^{\star}$Institut Universitaire de France\\
$^{\ddag}$IRISA, Universit\'e de Rennes, 35042 Rennes, France \\
}

\date{}

\maketitle


\begin{abstract}
This paper presents a simple generalization of causal consistency
suited to any object defined by a sequential specification.  As
causality is captured by a partial order on the set of operations
issued by the processes on shared objects (concurrent operations are
not ordered), it follows that causal consistency allows different
processes to have different views of each object history.

~ 
\end{abstract}

\section{Processes and Concurrent Objects}
Let us consider a set of $n$ sequential asynchronous processes 
$p_1$, ..., $p_n$, which cooperate by accessing shared objects. 
These objects are called {\it concurrent} objects. A main issue 
consists in defining the correct behavior of concurrent objects. 
Two classes of objects can be distinguished according to 
way they are specified.
\begin{itemize}
\vspace{-0.2cm}
\item 
The objects which can be defined by a sequential specification. 
Roughly speaking, this class of objects includes all the objects 
encountered in sequential computing (e.g., queue, stack, set, 
dictionary, graph). Different tools can be used to
define their correct behavior (e.g., transition function, 
list of all the correct traces -histories-, pre and post-conditions, etc.). 

It is usually assumed that the operations accessing these objects are 
{\it total},  which means that, whatever the current state of the object, 
an operation always returns a result. 

As an example, let us consider a bounded stack.  A $\pop()$ operation
returns a value if the stack is not empty, and returns the value
$\mathit{\vide}$ if it is empty.  A $\push(v)$ operation returns the
value $\mathit{\top}$ if the stack is full, and returns
$\mathit{\done}$ otherwise ($v$ was then added to the stack).  A
simpler example is a read/write register, where a read operation
always returns a value, and a write operation always returns ${\done}$.
\vspace{-0.2cm}
\item 
The objects which cannot be defined by a sequential specification. 
Example of such objects are Rendezvous objects or Non-blocking atomic commit
objects~\cite{G78}. These objects require processes to wait each other, 
and their correct behavior cannot be captured by  sequences of operations 
applied to them.  
\end{itemize}
In the following we consider objects defined by a sequential 
specification.

\section{Strong Consistency Conditions}

Strong consistency conditions are {\it natural} 
(and consequently easy to understand and use) in the sense that 
they require each object to appear as if it has been accessed sequentially. 
In a failure-free context, this can be easily obtained by using 
mutual exclusion locks bracketing the invocation of each operation.

\paragraph{Atomicity/Linearizability}
The most known and used consistency condition is {\it atomicity}, also 
called {\it linearizability}\footnote{Atomicity was formally defined 
in~\cite{L86,M86} for basic read/write objects. It was then generalized 
to any object defined by a sequential specification in~\cite{HW90}.
We consider these terms as synonyms in the following.}.
It requires that each object appears as if it was accessed sequentially, 
this sequence of operations belonging to the specification of the object, 
and complying with the real-time order of their occurrences.

\paragraph{Sequential consistency}
This consistency condition, introduced in~\cite{L79}, is similar to, but 
weaker than, linearizability, namely, it does not require the sequence 
of operations to comply with  real-time order. 

\begin{figure}[htb]
\centering{
\scalebox{0.4}{\input{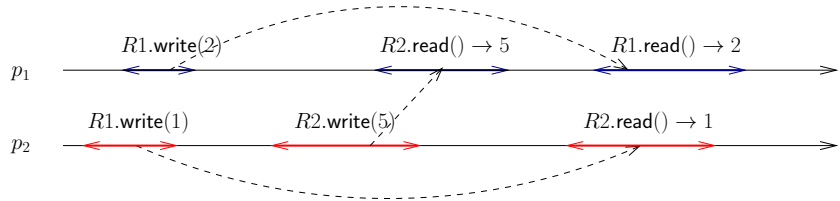}}
\caption{A  sequentially  consistent computation (which is not atomic)}
\label{fig:example-seq-consistency}
}
\end{figure}

Figure~\ref{fig:example-seq-consistency} presents an example of a
sequentially  consistent computation (which is not atomic) 
involving two read/write registers $R1$ and $R2$, accessed by two processes 
$p_1$ and $p_2$.  
The dashed arrows define the {\it causality} relation linking the read and 
write operations on each object.  It is easy to see that the  sequence of 
operations made up of  all the operations issued by $p_2$, followed by 
all the operations  issued by $p_1$, satisfies the definition of sequential 
consistency.

\paragraph{Implementing a strong consistency condition in an
asynchronous  message-passing  system}
Shared memories usually provide processes with  objects
built on top of basic atomic read/write objects or more 
sophisticated objects accessed by atomic operations such as Test\&Set or 
Compare\&Swap~\cite{H91,HS08,R13-b,T06}.
This is no longer the case in message-passing systems where all the objects 
(except communication channels) have to be built from scratch~\cite{AW04,R13}. 

Implementations of sequentially consistent objects and atomic objects in 
failure-free message-passing systems can be found 
in~\cite{AW94,AW04,CFJMR10,R02,R13}.
These implementations rest on a mechanism which allows a total order on all 
operations to be built. This can be done by  a central server, or  a broadcast 
operation delivering messages in the same order at all the processes. 
Such an operation is usually called {\it total order broadcast} (TO-broadcast) 
or {\it atomic broadcast}.  It is shown in~\cite{R02} that,
from an implementation point of view, sequential 
consistency can be seen as a form of lazy linearizability. 
The ``compositional'' power of sequential consistency is addressed 
in~\cite{EH16,PPMJ16}.

Implementations of a strong consistency condition
(such as atomicity)  in failure-prone message-passing
systems is more difficult. More precisely, except for a few objects
including read/write registers (which can be built only in systems
where, in each execution, a majority of processes do not
crash~\cite{ABD95}), it is impossible to implement an atomic object in
the presence of asynchrony and process crashes~\cite{FLP85}. Systems
have to be enriched with additional computing power (such as
randomization or failure detectors) to be able to implement objects
defined by a  strong consistency condition.

\section{Causal Consistency on Read/Write Objects (Causal Memory)}
\label {sec:Causal-cons-for-RW}

\paragraph{Causality-based consistency condition}
A causal memory is a set of read/write objects satisfying a
consistency property weaker that atomicity or sequential consistency.
This notion was introduced in~\cite{ANBHK95}.  It relies on a notion
of {\it causality} similar to the one introduced in~\cite{L78} for
message-passing systems.

The main difference between
causal memory and the previous strong consistency conditions
lies in the fact that  causality is captured by a partial order, 
which is trivially weaker than a total order. 
A total order-based consistency condition forces all the processes 
to see the same order on the object operations. Causality-based 
consistency does not. Each process can have its own view
of the execution, their  ''greatest common view'' being 
the causality partial order produced by the execution.   
Said differently,  an object defined by a strong consistency condition
is a {\it single-view} object, while an object defined by a causality-based 
consistency condition is a {\it multi-view} object (one view per process).

Another difference between a causality-based consistency condition 
and a strong consistency condition lies in the fact that 
a causality-based consistency condition copes naturally with process 
crashes and system partitioning.

\paragraph{Preliminary definitions}
As previously indicated, a causal memory is a set of read/write registers.
Its  semantics is based on the following preliminary definitions 
(from~\cite{ANBHK95,HW90}). 
To simplify the presentation and without loss of generality, we assume that 
(a) all the values written in a register are different, and (b)
each register has an initial value written by a fictitious write operation.  
\begin{itemize}
\vspace{-0.2cm}
\item A {\it local (execution) history}  $L_i$ of a process $p_i$ is the
sequence of read and write operations issued by this process.
If the operations $\op1$ and  $\op2$ belong to $L_i$ and 
$\op1$ appears before  $\op2$, we say ``$\op1$ precedes $\op2$ in $p_i$'s
process order''. This is denoted  $\op1 \stackrel{i~}\rightarrow\op2$.

\vspace{-0.2cm}
\item 
The {\it write-into relation} (denoted $\stackrel{wi}\rightarrow$)
captures the effect of write operations on the read operations.
Denoted $\stackrel{wi}\rightarrow$, it is defined as follows:
 $\op1 \stackrel{wi}\rightarrow\op2$ if
 $\op1$ is the write of a value  $v$ into a register $R$ and 
$\op2$ is a read operation of the register $R$ 
which returns the  value $v$. 
\vspace{-0.2cm}
\item An {\it execution history} $H$ is a partial order 
composed of one local history  per process, and a partial order,
denoted $\stackrel{po}\rightarrow$, defined as follows: 
$\op1 \stackrel{po}\rightarrow \op2$ if
\begin{itemize}
\vspace{-0.1cm}
\item 
$\op1, \op2\in L_i$ and  $\op1 \stackrel{i}\rightarrow \op2$
(process order),  or 
\vspace{-0.1cm}
\item 
$op1 \stackrel{wi}\rightarrow op2$ (write-into order), or
\vspace{-0.1cm}
\item 
$\exists ~\op3$ such that $\op1 \stackrel{po}\rightarrow \op3$
and  $\op3 \stackrel{po}\rightarrow \op2$ (transitivity).
\end{itemize}
\item 
Two operations not related by  $\stackrel{po}\rightarrow$ are said to be 
{\it independent} or {\it concurrent}.
\vspace{-0.2cm}
\item The projection of $H$ on a register $R$ (denoted $H|R$) 
is the partial order  $H$  from which are suppressed all the operations 
which are not on $R$. 
\vspace{-0.2cm}
\item A {\it serialization} $S$ of an execution history $H$ 
(whose partial order is  $\stackrel{po}\rightarrow$)
is a total order such that, if $\op1 \stackrel{po}\rightarrow \op2$, then
$\op1$ precedes $\op2$ in $S$. 
%
\end{itemize}

\paragraph{A remark on the partial order relation}
As we can see, the read-from relation mimics the causal send/receive 
relation associated with message-passing~\cite{L78}. The difference is 
that zero, one, or several reads can  be associated with the same write.
In both cases, the (write-into or message-passing) causality relation 
is a global property (shared by all processes) on which is built 
the consistency condition. It captures the effect of the environment 
on the computation (inter-process asynchrony), while process orders 
capture the execution of the algorithms locally executed by each process.

\paragraph{Causal memory}
Let $H_{i+w}$ be the partial order  $\stackrel{po}\rightarrow$,
from which all the read operations not issued by $p_i$ are suppressed. 
As defined in~\cite{ANBHK95}, an execution  history  $H$ is {\it causal} if,
for each process $p_i$, there is a serialization $S_i$ of  $H_{i+w}$
in which each read from a register  $R$ returns the value written in 
$R$ by the most recent preceding write in $R$. 

This means that, from the point of view of each process $p_i$, 
taken independently from the other processes, 
each register behaves as defined by its sequential specification. 
It is important to see, that different processes can have different views of 
a same register, each corresponding to a particular 
serialization of the  partial order  $\stackrel{po}\rightarrow$
from which the read operations  by the other processes have been eliminated.

\begin{figure}[htb]
\centering{
\scalebox{0.4}{\input{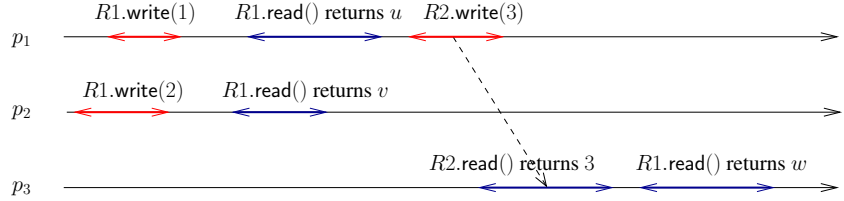}}
\caption{Example of an execution of a causal read/write memory}
\label{fig:causal-memory}
}
\end{figure}

An example of a causal memory execution is depicted in 
Figure~\ref{fig:causal-memory}.
Only one write-into pair is indicated  (dashed arrow). 
As $R1.\wwrite(1)$ and  $R1.\wwrite(2)$ are independent, each of the operations 
$R1.\rread()$ by $p_2$ and $p_3$ can return any value, i.e., $u,v\in \{1,2\}$. 
For the same reason, and despite the write-into pair on the register $R2$
involving $p_1$ and $p_3$, 
the operation $R1.\rread()$ issued by $p_3$ can return $w\in \{1,2\}$. 
This shows that different processes can obtain different ``views'' of 
the same causal memory execution. 
Once a read returned a value, a new  write-into pair is established. 

Implementations of a causal read/write memory (e.g.,~\cite{ART98})
rest on an underlying communication 
algorithm providing causal message delivery~\cite{BJ87,RST91}. 
It is shown in~\cite{ANBHK95,RS95} that, in executions 
that are data race-free or concurrent write-free,  a causal memory 
behaves as a sequentially consistent read/write memory.

\section{Causal Consistency for any Object}

\paragraph{The problem}
Albeit it was introduced more than 20 years ago, it appears that, 
when looking at the literature, causal consistency has been 
defined and investigated  only for read/write objects
(the only exception we are aware of is~\cite{PMJ16}). 
This seems to be due to the strong resemblance between read/write 
operations and send/receive operations. Hence, the question: 
Is it possible to generalize causal consistency to any object defined 
by a sequential specification? This section answers positively this question.

\paragraph{Preliminary definitions}
The notations and terminology are the same as in the previous section, but now
the operations are  operations on any object $O$ of a set of objects
$\cal O$, each defined by a sequential specification.

Considering a set of local histories and a partial order
$\stackrel{po}\rightarrow$ on their operations, let
$\assign_i(\stackrel{po}\rightarrow)$ denote the partial order
$\stackrel{po}\rightarrow$, in which, for each operation $\op()$ not
issued by $p_i$, the returned value $v$ is replaced by a value $v'$,
possibly different from $v$, the only constraint being that $v$ and
$v'$ belong to the same domain (as defined by the corresponding operation
$\op()$).  Let us notice that
$\assign_i(\stackrel{po}\rightarrow)$ is not allowed to modify the
values returned by the operations issued by $p_i$. Moreover, 
according to the domain of values returned by the operations, a lot of
different assignments can be associated with each process $p_i$.


Given a partial order  $\stackrel{po}\rightarrow$, and an operation $\op$, 
the {\it causal past} of $\op$ with respect to  $\stackrel{po}\rightarrow$
is the set of operations $\{\op'~|~\op' \stackrel{po}\rightarrow \op\}$.
A serialization $S_i$ of a partial order $\stackrel{po}\rightarrow$
is  said to be {\it causal past-constrained}  if it is 
such that, for any operation $\op$  issued by $p_i$, only the operations
of the causal past of $\op$ appear before $\op$.

\paragraph{Causal consistency for any object}
Let $H=\langle L_1,\ldots,L_n\rangle$ be a set of $n$ local histories 
(one per process) which access a set $\cal O$ of concurrent objects,
each defined by a sequential specification. 
$H$ is {\it causally consistent} if there is a partial order
 $\stackrel{po}\rightarrow$ on the operations of $H$ such that
for any process $p_i$:
\begin{itemize}
\vspace{-0.2cm}
\item
     $(\op1 \stackrel{i}\rightarrow \op2)
      \Rightarrow  (\op1 \stackrel{po}\rightarrow \op2)$, and
\vspace{-0.2cm}
\item 
      $\exists$ an assignment $\assign_i$ and  
      a causal past-constrained serialization  
      $S_i$ of  $\assign_i(\stackrel{po}\rightarrow)$ such that, 
      $\forall~O\in {\cal O}$, 
      $S_i|O$ belongs to the sequential specification of $O$.
\end{itemize}

The first requirement states that the partial order 
$\stackrel{po}\rightarrow$ must respect all process orders. The second
requirement states that, as far as each process $p_i$ is concerned,
the local view (of $\stackrel{po}\rightarrow$) it obtains is a total
order (serialization $S_i$) that, according to some value assignment, 
satisfies the sequential specification of each object $O$.\footnote{This 
definition is slightly stronger than the definition proposed in~\cite{PMJ16}. 
Namely, in addition to the introduction of the assignment notion, 
the definition introduced above adds the constraint that, if an operation 
$\op$ precedes an operation $\op'$ in the process order, then the 
serialization required for $\op$ must be a prefix of the 
serialization required for $\op'$. On the other hand, it describes precisely 
the level of consistency achieved by  Algorithm~\ref{algo:co-algorithm}
presented below.}


Let us remark that the assignments $\assign_i()$ and  $\assign_j()$
associated with $p_i$ and $p_j$, respectively, may provide  different
returned values in $S_i$ and $S_j$ for the same operation.
Each of them represents the local view of the corresponding process, 
which is causally consistent with respect to the global computation 
as captured by  the relation $\stackrel{po}\rightarrow$.

\paragraph{When the objects are read/write registers}
The definition of a causal memory stated in Section~\ref{sec:Causal-cons-for-RW}
is a particular instance of the previous definition. 
More precisely, given a process $p_i$, the assignment $\assign_i$
allows an appropriate value to be associated with every read not 
issued by $p_i$. Hence, there is a (local to $p_i$) assignment of values 
such that, in $S_i$, any read operation  returns the last written value. 
In a different, but equivalent way, the definition of a causal read/write 
memory given in~\cite{ANBHK95} eliminates from $S_i$ the read operations 
not issued  by $p_i$. 

While such operation eliminations are  possible for read/write objects, 
they are no longer possible when one wants to extend causal consistency to 
any object defined by a sequential specification. 
This come from the observation that, while a write operation resets 
``entirely'' the value of the object,  ``update'' operations on more 
sophisticated objects defined by a sequential specification (such as 
the operations $\push()$ and $\pop()$ on a  stack for example),  
do not  reset ``entirely'' the value of the object. The memory  of such 
objects has a richer structure than the one of a basic read/write object.

\paragraph{An example}
As an example illustrating the previous general definition of a
causally consistent object, let us consider three processes $p_1$
$p_2$ and $p_3$, whose accesses to a shared unbounded stack are
captured by the following local histories $L_1$, $L_2$, and $L_3$.  In
these histories, the notation $\op_i(a)r$ denotes 
the operation $\op()$ issued by $p_i$, with the input parameter $a$, 
and whose returned value is $r$.

\begin{itemize}
\vspace{-0.2cm}
\item $L_1~=~ \push_1(a)\done,~ \push_1(c)\done,\pop_1()$$c.$
\vspace{-0.2cm}
\item $L_2~=~ \pop_2()a,~ \push_2(b)\done,\pop_2()$$b.$
\vspace{-0.2cm}
\item $L_3~=~ \pop_3()a,~\pop_3()b.$
\end{itemize}
  
\begin{figure}[htb]
\centering{
\scalebox{0.4}{\input{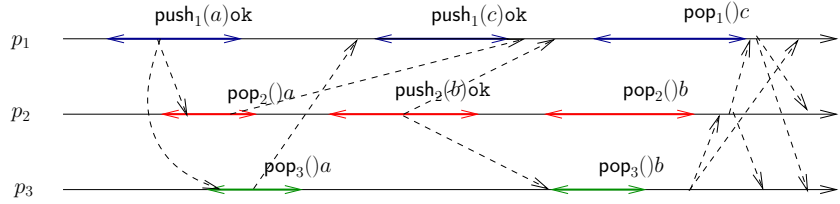}}
\caption{Example of a partial order on the operations issued on a stack}
\label{fig:example-cc-stack}
}
\end{figure}

Hence, the question: Is $H=\langle L_1,L_2,L_3 \rangle$ causally consistent?
We show that the answer is ``yes''. To this end we need first to build 
a partial order $\stackrel{po}\rightarrow$ respecting the three local process 
orders. Such a partial order is depicted in Figure~\ref{fig:example-cc-stack}, 
where process orders are implicit, and the inter-process causal relation is
indicated with dashed arrows (let us remind that this relation 
captures the effect of the environment --asynchrony-- on the computation). 

The second step consists in building three serializations respecting
$\stackrel{po}\rightarrow$, $S_1$ for $p_1$, $S_2$ for $p_2$, and
$S_3$ for $p_3$, such that, for each process $p_i$, there is an
assignment of values returned by the operations $\pop()$ ($\assign_i()$), 
from which it is possible to obtain a serialization $S_i$ belonging to the 
specification of the stack. Such assignments/serializations are given below.

\begin{itemize}
\vspace{-0.2cm}
\item 
$S_1~=~$
$\push_1(a)\done,~\pop_3$$()a,~\push_1(c)\done,~\pop_2()\vide,
             ~\push_2($$b$$)\done,~\pop_1()$$c$$,~\pop_2()$$b$$,~\pop_3()\vide.$ 
\vspace{-0.2cm}
\item 
$S_2~=~$
$\push_1(a)\done,~ \pop_2()$$a$$,~\push_2(b)\done,~\pop_2()$$b$$,~\pop_3()\vide,
~\pop_3()\vide, \push_1($$c$$)\done, ~\pop_1()$$c$$.$
\vspace{-0.2cm}
\item 
$S_3~=$
$\push_1(a)\done,~\pop_3()$$a$$,~\pop_2()\vide,~\push_2(b)\done,~\pop_3()$$b$$,
~\pop_2()\vide,~\push_1($$c$$)\done,~\pop_1()$$c$$.$
\end{itemize}

The local view of the stack of  each process $p_i$ is  constrained  
only by the causal order depicted in Figure~\ref{fig:example-cc-stack},  
and also depends on the way it orders concurrent operations. 
As far as $p_2$ is concerned we have the following,
captured by its serialization/assignment  $S_2$. (The serializations $S_1$ 
and $S_3$ are built similarly.) We have considered short local histories, 
which could be prolonged by adding other operations. 
As depicted in the figure, due to the last causality (dashed) arrows, 
those operations would have all the operations in $L_1\cup L_2\cup L_3$
in their causal past.

\begin{enumerate}
\vspace{-0.2cm}
\item Process $p_2$ sees first 
$\push_1(a)\done$, and consequently (at the implementation level) updates 
accordingly its local representation of the stack. 
\vspace{-0.2cm}
\item 
Then, $p_2$ sees its own invocation of $\pop_2()$ which returns it the 
value $a$. 
\vspace{-0.2cm}
\item 
Then, $p_2$ sees its own $\push_2(b)$ and $\pop_2()$ operations; 
$\pop_2()$ returns consequently $b$. 
\vspace{-0.2cm}
\item
Finally $p_2$ becomes aware of the two operations  $\pop_3()$ 
issued by $p_3$, and the operations $\push_1(c)$ and $\pop_1()$ 
issued by $p_1$. To have a consistent view of the stack, 
it considers the assignment of returned values that 
assigns the value  $\vide$ to the two operations $\pop_3()$, and 
the value  $c$ to the  operations $\pop_1()$. In this way, $p_2$ 
has a consistent view of the stack, i.e.,  a view which complies with 
the sequential  specification of a stack.
\end{enumerate}

\paragraph{A simple implementation}
A very simple implementation of causal consistency for any set of objects 
defined by sequential specifications can be be obtained from any underlying 
algorithm implementing  causal broadcast message delivery~\cite{BJ87,RST91}. 
Such a layered implementation, which considers deterministic objects, 
is described in Figure~\ref{algo:co-algorithm}\footnote{Interestingly, 
the replacement in this algorithm of the underlying message causal order 
broadcast  by a message total order broadcast, implements linearizability.}. 
Let  ``${\cob}$ {\sc msg}$(a)$'' denote the causal broadcast of 
a message tagged {\sc msg}$(a)$ carrying the value $a$.  The associated 
causal reception at any process is denoted ``co-delivery''. ''?''
denotes a control value unknown by the processes at the application level. 

\begin{algorithm}[h!]
\centering{\fbox{
\begin{minipage}[t]{150mm}
\footnotesize 
\renewcommand{\baselinestretch}{2.5}
\begin{tabbing}
aaaaa\=aaa\=aaaaa\=aaaaaa\=\kill

{\bf when} $p_i$ {\bf invokes}  $O.\op(param)$ {\bf do}\\

\line{CO-01} \>  $result_i \leftarrow~?$;\\

\line{CO-02} \>  ${\cob}$ {\sc operation}$(i,O,\op(param))$;\\ 

\line{CO-03} \>  $\wait$   $(result_i\neq~ ?)$;\\

\line{CO-04} \>  $\return$ $(result_i)$.\\~\\

{\bf when} {\sc operation}$(j,O,\op(param))$ {\bf is} 
${\sf co\mbox{-}delivered}$ {\bf do}\\

\line{CO-05} \> $\langle r,state_i[O]\rangle \leftarrow 
                                \delta_O(state_i[O],\op(param))$;\\

\line{CO-06} \>  {\bf if} $(j=i)$   
               {\bf then}  $result_i \leftarrow r$  {\bf end if}.

\end{tabbing}
\end{minipage}
}
\caption{An implementation of causal order (code for $p_i$)}
\label{algo:co-algorithm}
}
\end{algorithm}

Each object $O$ is defined by a transition function $\delta_O()$, which takes 
as input  parameter the current state of $O$ and the operation $\op(param)$
applied to $O$. It returns a pair $\langle r,new\_state \rangle$, where $r$ is 
the value returned by $\op(param)$, and $new\_state$ is the new state of $O$. 
Each process $p_i$ maintains a local representation of each object $O$, 
denoted $state_i[O]$.

When a process $p_i$ invokes an operation  $\op(param)$ on an object $O$, it 
co-broadcasts the message  {\sc operation}$(i,O,\op(param))$, which 
is co-delivered to each process (i.e., according to causal message order). 
Then, $p_i$ waits until this message is locally processed. 
When this occurs, it returns the result of the operation. 

When a process $p_i$ co-delivers a message {\sc operation}$(j,O,\op(param))$, 
it updates accordingly its local representation of the object $O$.  If $p_i$ 
is the invoking process, it additionally locally returns the result of the 
operation.


\section{Conclusion}
This research note has introduced the notion of causal consistency for 
any object defined by a sequential specification. This definition
boils down to causal memory when the objects are read/write registers. 

The important point in causal consistency lies in the fact that each process 
has its own view  of the objects,  and all these views agree on the partial 
order on the  operations but not necessarily on their results. 
More explicitly,  while each process has a view of each object, which  
locally satisfies its object specification, two processes may disagree on the 
value returned by some operations. This seems to be the ``process-to-process 
inconsistency cost'' that must be paid when weakening consistency 
by considering a partial order instead of a total order. 
On another side, differently from strong consistency conditions, 
causal consistency copes naturally with partitioning and process crashes. 

\section*{Acknowledgments}
This work has been partially supported by the  Franco-German
DFG-ANR Project 40300781 DISCMAT (devoted to connections between
mathematics and distributed computing), and the French ANR project DESCARTES 
(devoted to layered and modular structures in distributed computing). 


\end{document}